\documentclass[journal=nalefd,manuscript=letter,layout=twocolumn]{achemso}

\usepackage{chemformula} 
\usepackage[T1]{fontenc} 
\usepackage{float}
\usepackage{hyperref}

\author{Xinyue Zhu}
\author{Yu Xie}
\email{xieyu@cumt.edu.cn}
\author{Yifei Hao}
\author{Fei Gao}
\author{Yue Li}
\author{Junting Zhang}
\email{juntingzhang@cumt.edu.cn}
\affiliation{School of Materials Science and Physics, China University of Mining and Technology, Xuzhou 221116, China}

\title{Topological phase transition driven by in-plane spin rotation}

\keywords{Topological phase transition, Chern insulator, Spin reorientation, Berry curvature, Two-dimensional kagome magnet}

\usepackage{stfloats}
\begin{document}

\begin{abstract}
The intrinsic coupling between magnetism and nontrivial band topology in magnetic topological insulators makes external magnetic fields a powerful tool for manipulating topological states. However, conventional magnetic control mechanisms, such as driving magnetic phase transitions or fully reversing magnetization, typically demand large magnetic fields and lack continuous tunability. Here, we establish a symmetry framework for the reversible switching of topological states via continuous in-plane spin rotation, governed by magnetic point group constraints on the Berry curvature distribution. Using a two-dimensional kagome ferromagnetic Chern insulator as a prototype, we demonstrate that a $60^\circ$ in-plane magnetization rotation reverses the sign of the Chern number, transitioning through a topologically trivial state. Crucially, micromagnetic simulations confirm that this spin-reorientation-driven switching operates under exceptionally small magnetic fields and on ultrafast timescales. This work provides a highly efficient, low-energy paradigm for the manipulation of topological states.
\end{abstract}

\hspace*{\fill} \\
Magnetic topological insulators (MTIs), characterized by the intrinsic coupling between magnetic order and nontrivial band topology, manifest a diverse array of exotic quantum states, including the quantum anomalous Hall effect, axion insulator states, and topological magnetoelectric effects. \cite{Deng_quantum_2020,Tokura_magnetic_2019,Liu_robust_2020,Xiao_realization_2018,Zhang_topological_2019,Wang2023} The hallmark of MTIs lies in their topologically protected edge states, which exhibit robust spin-momentum locking and dissipationless transport that hold great promise for applications in next-generation ultralow-power spintronics and topological quantum computing. \cite{He_topological_2022,Hasan_colloquium_2010,Pesin_spintronics_2012,Qi_topological_2011,Li_electrical_2014,Shiomi_spin_2014} Currently, research on MTIs has predominantly focused on the MnBi$_2$Te$_4$ material family, which exhibits a rich landscape of diverse topological physical phenomena and highly tunable topological properties.\cite{Chen_intrinsic_2019,Hao_gapless_2019,Li_intrinsic_2019,Otrokov_prediction_2019} However, its relatively low magnetic transition temperature remains a critical bottleneck, impeding its integration into practical electronic devices. \cite{Yan_crystal_2019} Consequently, the search for MTIs with high magnetic transition temperatures remains a key challenge in this field.

Precisely manipulating the topological electronic states of MTIs via external fields is essential for their application in devices. Optically, ultrafast mid-infrared pulses can effectively tune the Dirac mass gap of MnBi$_2$Te$_4$, \cite{Bielinski_floquet_2025} enabling dynamic control over its topological states.\cite{Oka_floquet_2019} Electrically, gating or vertical displacement fields are typically employed to modulate carrier concentrations and break symmetries.\cite{Yang_electrically_2025,Zhang_design_2020,Zhang_coexistence_2022,Wang_magnetoelectric_2023,li2025generation,Xie_manipulation_2025} For example, displacement fields can induce a Chern insulator phase and tune the Chern number in moiré superlattices, \cite{Chen_tunable_2020} while gate-driven doping is theoretically predicted to achieve similar control in orbital Chern insulators. \cite{Zhu_voltage_2020} Nevertheless, magnetic field control remains the most prevalent strategy due to the intrinsic coupling between band topology and magnetic order parameters.\cite{Gong2019,Lee2020,Rienks2019} Applying a magnetic field can drive an antiferromagnetic-to-ferromagnetic transition in MnBi$_2$Te$_4$, triggering a topological phase transition from an axion to a Chern insulator state. \cite{Liu_robust_2020,Zhang_topological_2019} Furthermore, a $180^\circ$ magnetization reversal induced by magnetic field switches the sign of the Chern number.\cite{Chang_experimental_2013,Yu_quantized_2010} However, such magnetic control typically requires a large magnetic field and lacks continuous tunability, highlighting a critical bottleneck in current magnetic field control schemes.

Exploiting spin-orientation-driven topological phase transitions offers a compelling alternative to conventional magnetic control. Theoretically, the spin orientation determines the magnetic point group (MPG) symmetries, which in turn govern the momentum-space distribution of the Berry curvature.\cite{Xiang_prediction_2015,Liu_Chern_2023,ATang_Comprehensive_2019} Therefore, altering the spin orientation can induce a reconstruction of the Berry curvature, which may cause a topological phase transition in the system. For instance, switching magnetization from out-of-plane to in-plane can annihilate topological electronic states in specific systems. \cite{Liu_Intrinsic_2018} However, this process typically demands overcoming a large magnetic anisotropy energy barrier. Recently, it was demonstrated that a non-$180^\circ$ in-plane rotation of the magnetization can trigger a sign reversal of the Chern number in certain MTIs. \cite{Li_chern_2022, Liu_inplane_2013} Nevertheless, a systematic symmetry framework to elucidate this topological phase transition driven by in-plane spin rotation remains elusive.

In this work, we establish a symmetry framework governing topological phase transitions driven by in-plane spin rotation in 2D Chern insulators, rooted in the constraints of MPG symmetry on the distribution of Berry curvature. This symmetry-based framework transcends the limitations of previous material-specific studies, establishing a universal design principle for screening candidate materials that host similar topological phase transitions. By combining tight-binding model, first-principles calculations, and micromagnetic simulations, we elucidate the microscopic origin of this topological phase transitions and demonstrate robust topological manipulation using a small magnetic field. Specifically, we introduce the 2D kagome magnet $\text{Ni}_3\text{Sb}_2\text{O}_8$ as a candidate material system to demonstrate its spin-orientation-driven topological phase transitions. We reveal that a $60^\circ$ in-plane magnetization rotation is sufficient to reverse the sign of the Chern number via an intermediate trivial phase, requiring only a minimal switching field on an ultrafast picosecond timescale. Compared to conventional magnetic control, this in-plane rotation strategy offers a significantly faster and low-energy paradigm for manipulating topological states.

\begin{figure*}[t]
\centering
\includegraphics[width=0.95\textwidth]{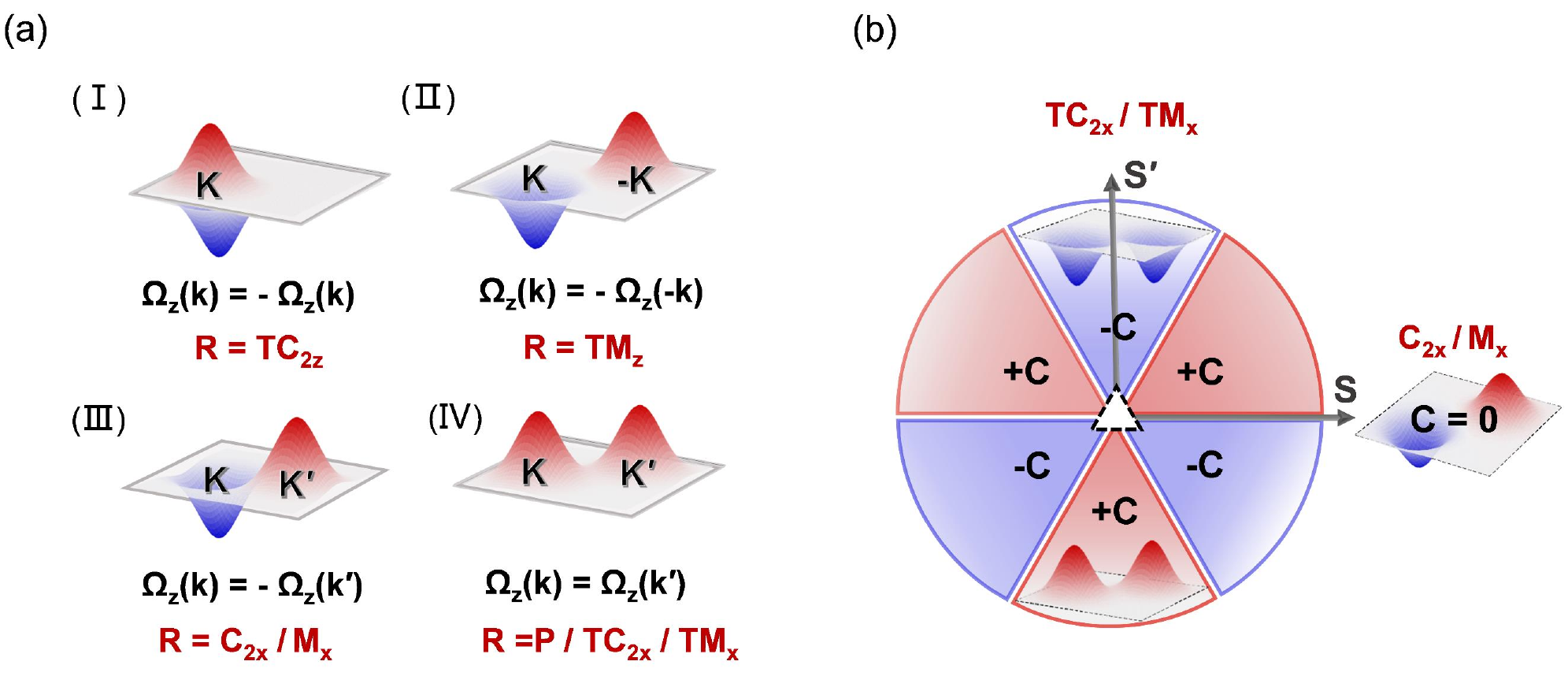}
\caption{\label{Figure 1} (a) Constraints of magnetic symmetry operations on the distribution of Berry curvature $\Omega(\mathbf{k})$ in 2D momentum space. (b) Phase diagram mapping the Chern number $C$ as a function of the in-plane spin orientation in a centrosymmetric $\bar{3}m$ system, with the twofold axis aligned along the $x$ axis.}
\end{figure*}

The topological phase of the system is fundamentally determined by the momentum-space distribution of the Berry curvature, $\Omega_z(\mathbf{k})$, whose integration over the first Brillouin zone defines the Chern number $C$:\cite{Thouless_quantized_1982,Xiao_berry_2010}
\begin{equation}
    C = \frac{1}{2\pi} \int_{\text{BZ}} \Omega_{z}(\mathbf{k}) \, d^2\mathbf{k}
\end{equation}
As a pseudo-scalar in a 2D system, $\Omega_z(\mathbf{k})$ is strictly constrained by the MPG symmetry. Specifically, it is invariant under inversion ($P$) but reverses sign under time reversal ($T$).\cite{Liu_Chern_2023,Liu_Ferroelectrically_2025} For an in-plane ferromagnetic system, the allowed symmetry operations consist of $R \in \{E, P, TC_{2z}, TM_z, C_{2x}, M_x, TC_{2x}, TM_x\}$, which impose four distinct constraints on the Berry curvature, as illustrated in Figure \ref{Figure 1}a. Specifically, (I) Retaining the $TC_{2z}$ operation enforces $\Omega_z(\mathbf{k}) \equiv 0$. (II) Retaining $TM_z$ yields an odd-parity distribution, $\Omega_z(\mathbf{k}) = -\Omega_z(-\mathbf{k})$, strictly enforcing $C = 0$. (III) The $C_{2x}$ and $M_x$ operators, occurring when magnetization aligns with an in-plane twofold axis or perpendicular to a vertical mirror plane, cause the symmetry-related momentum points to exhibit opposite Berry curvatures, $\Omega_z(\mathbf{k}) = -\Omega_z(\mathbf{k}')$. (IV) Conversely, operators such as $P$, $TC_{2x}$, and $TM_x$ enforce even-parity or symmetric distribution, $\Omega_z(\mathbf{k}) = \Omega_z(\mathbf{k}')$, thereby permitting a non-zero Chern number. Consequently, to host a non-trivial topological phase in an in-plane ferromagnet, the crystal structure must break both $C_{2z}$ and $M_z$ operations, restricting the operators along the out-of-plane principal axes to $C_{3z}$ or $PC_{3z}$. Crucially, to drive a topological phase transition via in-plane magnetization rotation, the crystal structure must inherently include either $C_{2x}$ or $M_x$ operations, as summarized in Table \ref{Table1}. Among the point groups satisfying this prerequisite, hexagonal systems (specifically 32, $3m$, and $\bar{3}m$) are highly preferred, given their negligible magnetic anisotropy barriers for in-plane spin rotation.

To illustrate the evolution of topological states with in-plane spin orientation, we use the centrosymmetric point group $\bar{3}m$ as a representative model. When the spin is oriented along a general in-plane direction, the simultaneous breaking of all rotational and mirror symmetries reduces the MPG to $\bar{1}$, thereby allowing the emergence of a non-zero Chern number. Aligning the spin perpendicular to the twofold axis transforms the MPG to $2'/m'$, which enforces an even-parity distribution of Berry curvature and sustains a nontrivial Chern number (see Figure \ref{Figure 1}b). Conversely, when the spin aligns parallel to the twofold axis, the MPG transitions to $2/m$, forcing the Chern number to zero. Furthermore, since the Berry curvature reverses sign under this twofold rotation, the resulting Chern numbers for $C_{2x}$ symmetry-related spin orientations must be opposite. These fundamental symmetry constraints establish a $120^\circ$  periodicity for the Chern number with respect to the in-plane spin azimuth, as illustrated in Figure \ref{Figure 1}b. Consequently, changing the in-plane spin direction provides a robust mechanism to trigger sign reversals of the Chern number and drive topological phase transitions from a Chern insulator to a trivial insulator state.

\begin{table*}[t]
\centering
\caption{\label{Table1} Classification of symmetry constraints for 2D point groups hosting in-plane ferromagnetic Chern insulators.}
\includegraphics[width=0.8\textwidth]{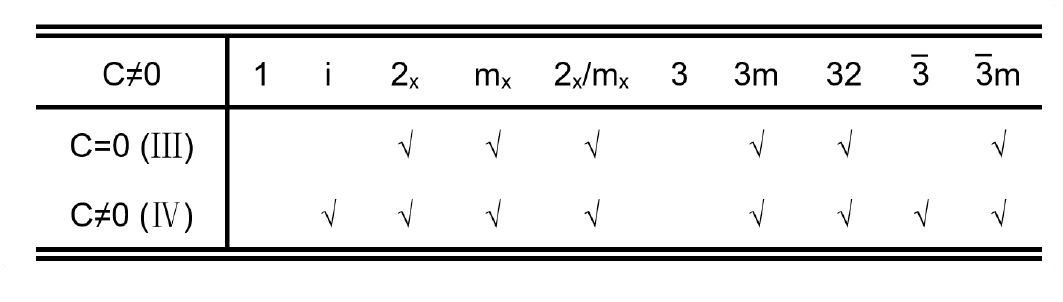}
\end{table*}

To elucidate the microscopic mechanism of topological phase transition driven by in-plane spin rotation, we construct a tight-binding model on a kagome lattice, with the Hamiltonian formulated as:
\begin{align}
    \mathbf{H} &= -t \sum_{\langle i\alpha, j\beta \rangle} c_{i\alpha}^{\dagger} c_{j\beta} + i\lambda_{I} \sum_{\langle i\alpha, j\beta \rangle} \mu_{ij} c_{i\alpha}^{\dagger} \sigma_{\alpha\beta}^{z} c_{j\beta} \nonumber \\
    &\quad + i\lambda_{R} \sum_{\langle i\alpha, j\beta \rangle} \nu_{ij} c_{i\alpha}^{\dagger}(\vec{\sigma} \cdot \vec{\eta}_{ij})_{\alpha\beta} c_{j\beta} \nonumber \\
    &\quad + B \sum_{i,\alpha,\beta} c_{i\alpha}^{\dagger} (\hat{m}(\theta, \varphi) \cdot \vec{\sigma})_{\alpha\beta} c_{i\beta} \label{eq:hamiltonian}
\end{align}
Here, $c_{i\alpha}^\dagger$ ($c_{j\beta}$) denotes the creation (annihilation) operator for an electron with spin $\alpha (\beta$) at site 
$i$ ($j$). $\vec{\sigma}$ represents the Pauli matrices, and the direction coefficients $\mu_{ij}$ and $\nu_{ij}$ take values of $\pm 1$. The first term represents the nearest-neighbor hopping between different sublattices. The second and third terms capture the intrinsic and Rashba spin-orbit coupling (SOC) characterized by strengths $\lambda_{\text{I}}$ and $\lambda_R$, respectively. The fourth term accounts for the magnetic exchange interaction, where $\hat{m}(\theta,\varphi)$ denotes the tunable magnetization orientation. The parameters utilized in this tight-binding model are extracted from first-principles calculations of the selected material $\mathrm{Ni}_{3}\mathrm{Sb}_{2}\mathrm{O}_{8}$. The topological phase diagram, calculated as a function of the intrinsic ($\lambda_I$) and Rashba ($\lambda_R$) SOC parameters, demonstrates that the Chern insulator phase is highly robust, emerging across a wide parameter space in the 2D kagome lattice (see Figure S1).

\begin{figure*}[t]
\centering
\includegraphics*[width=0.95\textwidth]{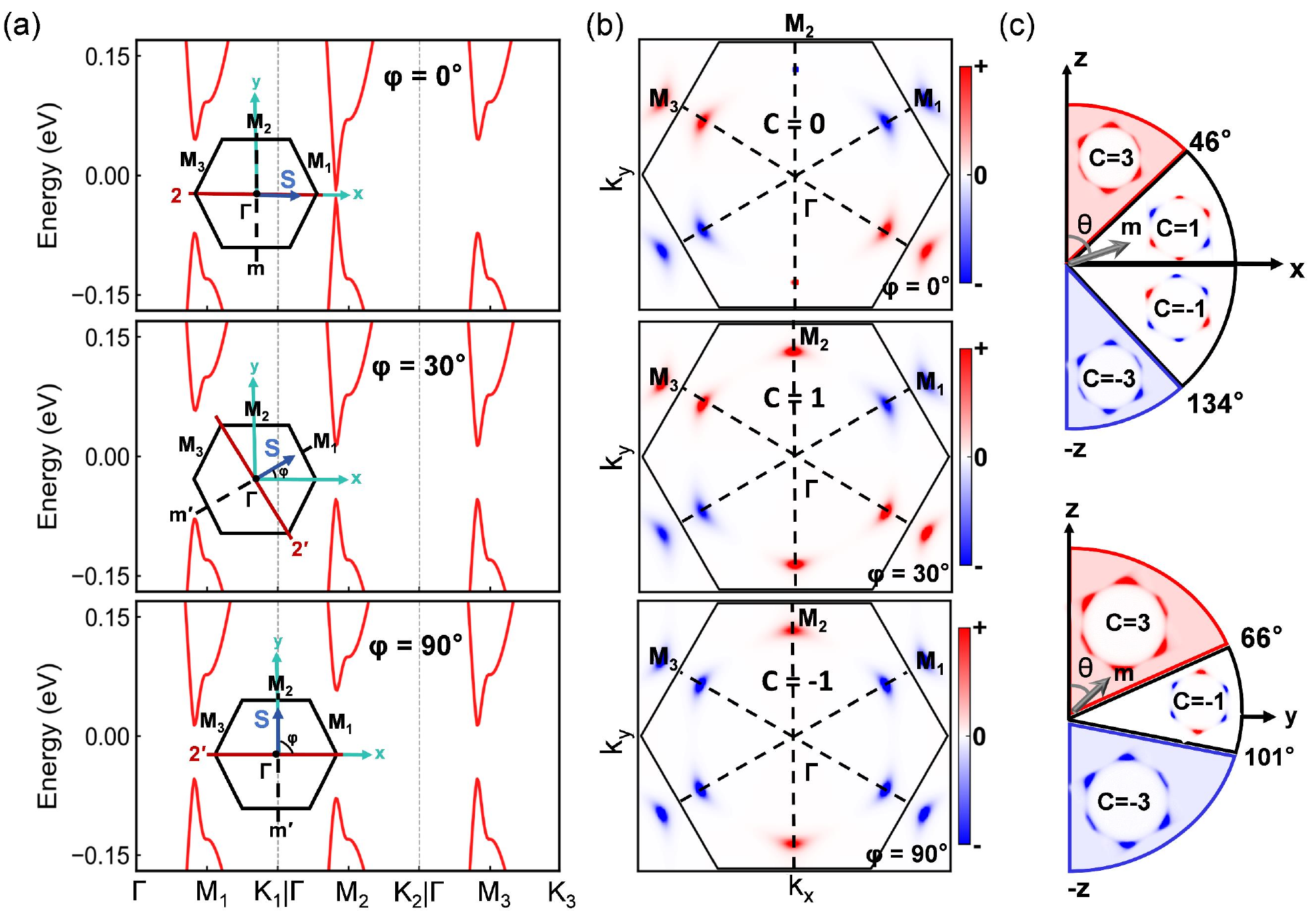}
\caption{\label{Figure 2} Spin-orientation-dependent topological properties derived from the tight-binding model. (a) Band structures along the high-symmetry $\Gamma$-$\mathrm{M}$ paths and (b) corresponding Berry curvature distributions for selected in-plane azimuthal angles ($\varphi =0^\circ$, $30^\circ$, $90^\circ$). The spin azimuthal angle is denoted as $\varphi$ (relative to the $x$ axis). The insets in (a) show the magnetic symmetry operations retained under the corresponding spin orientation. The dashed and solid lines represent the retained mirror planes and twofold axes, respectively. The tight-binding parameters are set as $t = 0.5$, $\lambda_{I} = 0.1$, $\lambda_{R} = 0.05$, and $B = 0.65$. (c) Topological phase diagrams illustrating the angular dependence of the Chern number for spin rotation within the $xz$ and $yz$ planes.}
\end{figure*}

Figure \ref{Figure 2}a shows the evolution of the band structure with the in-plane spin direction. At an azimuthal angle of $\varphi = 0^\circ$, the system retains a twofold rotation axis parallel to the spin direction and a orthogonal vertical mirror plane. This specific symmetry (Case III) triggers a gap closure along the $\Gamma$-M$_2$ path and results in an antisymmetric Berry curvature distribution with respect to this twofold axis, thus locking the system in a topologically trivial state (see Figure \ref{Figure 2}b). Rotating the spin to $\varphi=30^\circ$ reopens the band gap across all $\Gamma$-M paths and changes the magnetic symmetry to $2'_x/m'_x$ (Case IV). This symmetry enforces a symmetric Berry curvature distribution and drives the system into a Chern insulator phase ($C = 1$). A similar band topology and Berry curvature profile emerge at $\varphi =$ 90° due to identical symmetry constraints. However, these two states ($\varphi=30^\circ$ and $90^\circ$) host opposite Chern numbers, a direct consequence of being connected by the twofold rotation axis along $\varphi=60^\circ$. Therefore, continuous in-plane spin rotation provides a robust mechanism to switch the system between the $C = 1$ and $C = -1$ Chern insulator phases, transitioning via an intermediate topologically trivial state. 

Spin-reorientation-driven topological phase transition can also extends to out-of-plane configurations. Tilting the magnetization into the $xz$ or $yz$ planes (parallel and perpendicular to the twofold axis, respectively) drives the system into a high-Chern-number phase ($C = 3$) once a critical deviation angle is surpassed (see Figure \ref{Figure 2}c). Notably, the evolution of Chern number within the $xz$ plane exhibits a strictly antisymmetric dependence on the spin angle, a direct consequence of the symmetry constraints imposed by the twofold rotation about the $x$ axis.

\begin{figure*}[b]
\centering
\includegraphics[width=0.99\textwidth]{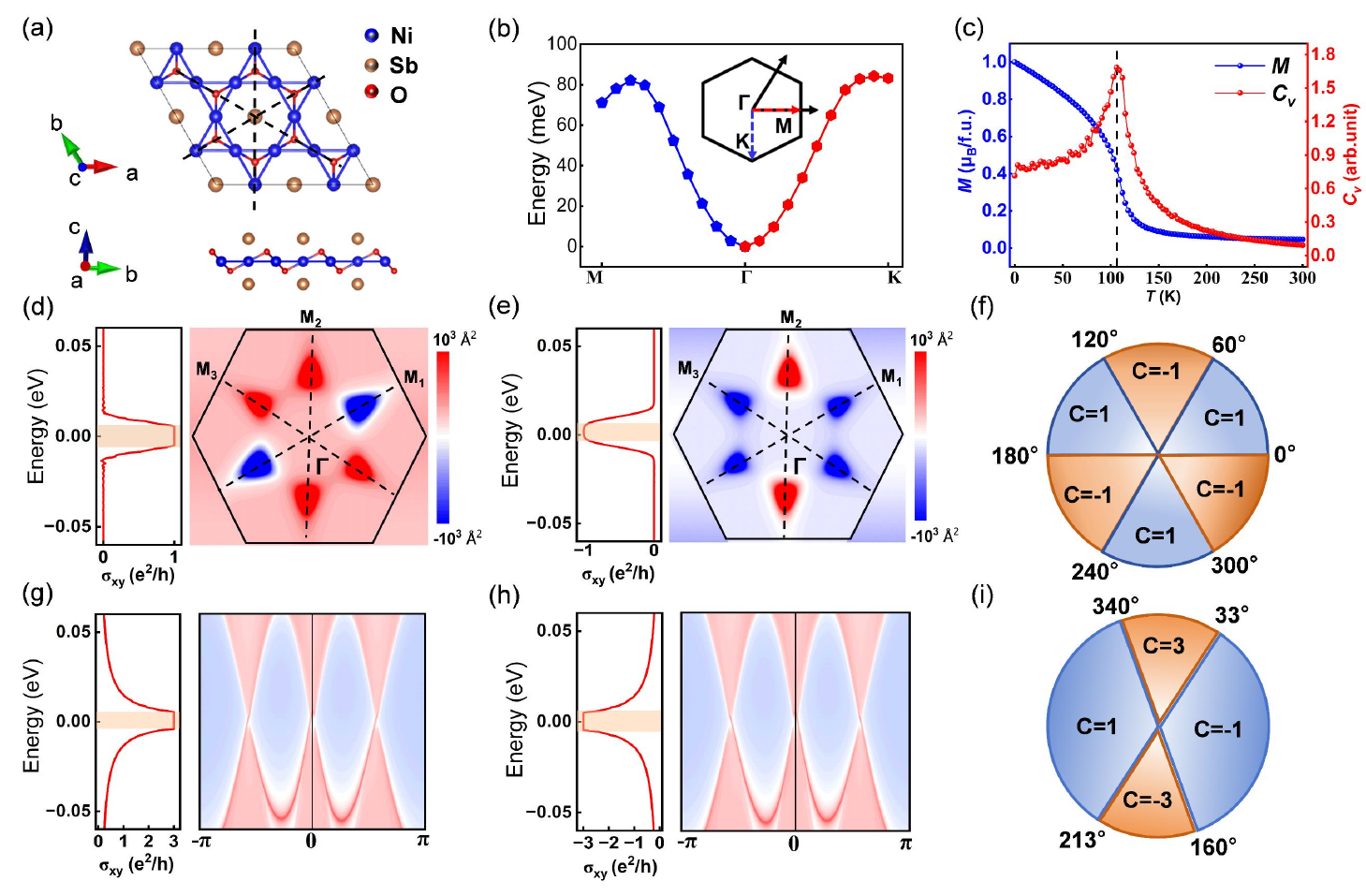}
\caption{\label{Figure 3} Topological properties of 2D kagome ferromagnet $\mathrm{Ni}_{3}\mathrm{Sb}_{2}\mathrm{O}_{8}$. (a) Top and side views of the crystal structure, where the black dashed lines denote the three vertical mirror planes. (b) Energy variation of spiral spin orders as a function of the wave vector $\mathbf{q}$ along high-symmetry paths. (c) Temperature dependence of the magnetization and specific heat. (d, e) Anomalous Hall conductivity and Berry curvature profiles for in-plane spin azimuthal angles $\varphi = 30^\circ$ and $\varphi = 90^\circ$, respectively. (f) Topological phase diagram illustrating the dependence of the Chern number on the spin azimuthal angle. (g, h) Anomalous Hall conductivity and chiral edge states corresponding to the out-of-plane spin directions $+z$ and $-z$, respectively. (i) Topological phase diagram illustrating the angular dependence of the Chern number as the spin rotates within the $yz$-plane.}
\end{figure*}

Next, we take the 2D kagome magnet $\mathrm{Ni}_{3}\mathrm{Sb}_{2}\mathrm{O}_{8}$ as a representative example to demonstrate the topological phase transition driven by spin rotation. Its crystal structure consists of edge-sharing oxygen octahedra wherein the central transition-metal ions form a prototypical kagome lattice \cite{yin2021kagome} (see Figure \ref{Figure 3}a). The system crystallizes in the centrosymmetric space group $P\bar{3}m1$, characterized by three vertical mirror planes along the $\Gamma$-$\mathrm{M}$ directions. We confirmed the dynamic and thermodynamic stability of $\mathrm{Ni}_{3}\mathrm{Sb}_{2}\mathrm{O}_{8}$ monolayer through phonon spectrum calculations and first-principles molecular dynamics simulations, respectively (see Figure S2).  

To determine the magnetic ground state, we calculated the energies of various collinear magnetic configurations (see Figure S3), indicating that the ferromagnetic order exhibits the lowest energy. To validate this ferromagnetic ground state, we calculated the energy variation of spiral spin orders with respect to the wave vector $\mathbf{q}$ using the generalized Bloch theorem.\cite{AAsandratskii1998noncollinear} The global energy minimum occurs at the $\Gamma$ point, confirming a ferromagnetic ground state (see Figure \ref{Figure 3}b). By employing the Green’s function method based on magnetic force theory,\cite{he2021tb2j,liechtenstein1987local} we extracted the magnetic exchange parameters, yielding $J_1 = 2.12$ meV and $J_2 = 9.71$ meV, indicating that the ferromagnetism is primarily governed by the next-nearest-neighbor magnetic exchange interaction. Additionally, Monte Carlo simulations further confirm the stability of ferromagnetic phase and estimate the Curie temperature ($T_{\mathrm{C}}$) of approximately $109\ \mathrm{K}$ (see Figure \ref{Figure 3}c). Magnetic anisotropy calculation indicate that this monolayer energetically prefers in-plane magnetization (see Figure S4).

First-principles calculations demonstrate that the topological phase of  $\mathrm{Ni}_{3}\mathrm{Sb}_{2}\mathrm{O}_{8}$ monolayer is the in-plane magnetization direction. The band structures calculated by density functional theory are in excellent agreement with the tight-binding model (see Figure S5). Spin orientation along the $x$-axis ($\varphi =0^\circ$) causes the system to enter a topologically trivial state (see Figure S6). Rotating the spin direction to $\varphi =30^\circ$ leads to a symmetric Berry curvature profile with respect to the $\Gamma$-$\mathrm{M}_1$ path (see Figure \ref{Figure 3}d). The emergence of gapless chiral edge states (see Figure S6) and the integer plateau in the anomalous Hall conductivity (Figure \ref{Figure 3}d) collectively indicate that the system transitions to a Chern insulator phase with $C = 1$, which exhibits a substantial global band gap of 30.6 meV (see Figure S7). Aligning the spin direction to $\varphi = 90^\circ$ redistributes the Berry curvature symmetrically about the $\Gamma$-$\mathrm{M}_2$ axis, driving the system into the $C = -1$ state (see Figure \ref{Figure 3}e and Figure S6). Microscopically, this topological transition originates from a band inversion driven by Rashba SOC. Specifically, the valence band maximum and conduction band minimum are predominantly composed of the Ni $d_{xz}$ and $d_{yz}$ orbitals, which undergo an exchange of their orbital characters upon a $60^\circ$ spin rotation (see Figure S8). Consequently, spin rotation within the plane yields a periodic evolution of the Chern number, where a $60^\circ$ rotation efficiently triggers a sign reversal of the Chern number, as illustrated in Figure \ref{Figure 3}f.

Topological phase transition driven by out-of-plane spin rotation is also demonstrated. When the spin is oriented along the out-of-plane direction (\emph{z} axis), a high-Chern-number topological state with $C = \pm 3$ emerge with a band gap of 29.6 meV, and reversing the spin direction results in a change in the sign of the Chern number (see Figure \ref{Figure 3}g and \ref{Figure 3}h). This high Chern number stems from the fact that the Berry curvature exhibits a symmetric distribution along all $\Gamma -\mathrm{M}$ paths (see Figure S9). Detailed evaluation of spin rotation within the $yz$ plane indicates that this phase transition occurs at the critical angles of $\theta_{c} = 33^\circ$ and $160^\circ$ (see Figure \ref{Figure 3}i). 

\begin{figure}[t]
\centering
\includegraphics*[width=0.5\textwidth]{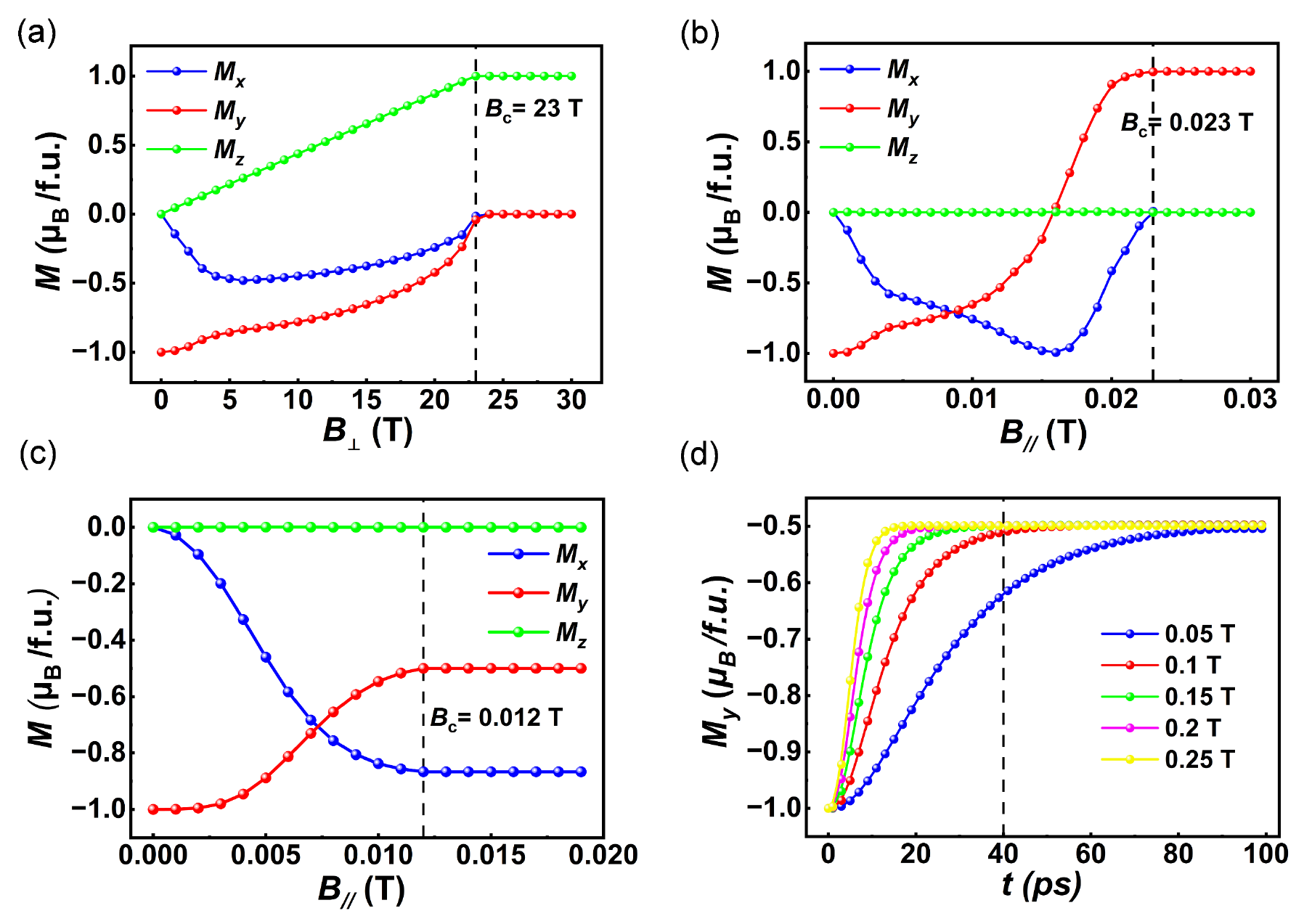}
\caption{\label{Figure 4} Micromagnetic simulations of magnetization switching dynamics. (a) Variation of magnetization components as functions of out-of-plane field ($B_\perp$). (b, c) Evolution of magnetization during  $180^\circ$ and $60^\circ$ in-plane rotation driven by an in-plane field ($B_\parallel$). (d) Time-resolved dynamics of the $M_y$ component under a $60^\circ$ in-plane magnetic field at different amplitudes.}
\end{figure}

To evaluate the practical feasibility of controlling these topological states via external magnetic fields, we simulated the magnetization dynamics using the Landau-Lifshitz-Gilbert (LLG) equation.\cite{gilbert2004phenomenological,AEvans_Atomistic_2014} In a finite-sized 2D kagome lattice, the initial magnetization aligns along the $-y$ axis, governed by the interplay of easy-plane magnetocrystalline anisotropy and shape anisotropy induced by the demagnetizing field. Consequently, the in-plane anisotropy induced by the demagnetizing field introduces an energy barrier during spin rotation within the $ab$ plane. Owing to the large magnetocrystalline anisotropy, forcing the magnetization out-of-plane demands a large critical field of $23\ \mathrm{T}$ (see Figure \ref{Figure 4}a). In contrast, a $180^\circ$ reversal of the in-plane magnetization can be readily achieved with a small applied field of $0.023\ \mathrm{T}$ (see Figure \ref{Figure 4}b). Crucially, driving a $60^\circ$ in-plane rotation, which is sufficient for topological switching, requires an even lower critical field of $0.012\ \mathrm{T}$, approximately half that of the full reversal (see Figure \ref{Figure 4}c). 

Beyond energy efficiency, we evaluated the dynamic performance of this magnetization switching scheme. The switching duration scales inversely with the applied field strength (see Figure \ref{Figure 4}d). Furthermore, a $60^\circ$ magnetization rotation occurs on a significantly faster timescale than a conventional $180^\circ$ reversal (see Figure S10). For instance, under a moderate applied magnetic field of $0.1\ \mathrm{T}$, the $60^\circ$ rotation completes in just $\sim 40\ \mathrm{ps}$, demonstrating an ultrafast dynamic response. Consequently, this scheme not only reduces the required magnetic field threshold but also offers a much faster response speed. 

In conclusion, we have proposed and validated an efficient strategy for controlling topological electronic states through modulating the in-plane spin orientation. We elucidate the symmetry conditions and  microscopic mechanisms governing this topological phase transitions driven by spin rotation in 2D Chern insulators. Using the 2D kagome magnet $\mathrm{Ni}_{3}\mathrm{Sb}_{2}\mathrm{O}_{8}$ as a prototype, we demonstrate that a mere $60^\circ$ in-plane spin rotation reversibly switch the Chern number via an intermediate trivial phase. Furthermore, tilting the magnetization out-of-plane induces high-Chern-number topological states with $C=\pm3$. Crucially, this non-$180^\circ$ switching scheme significantly reduces the required critical magnetic field while enabling an ultrafast dynamic response. Although the Curie temperature of the selected material remains well below room temperature, our proposed strategy is rooted in fundamental symmetry principles. Consequently, it can serve as a universal framework to guide the high-throughput screening of candidate materials hosting analogous topological phase transitions. This work pioneers a precise and energy-efficient avenue for the manipulation of band topology.

\begin{suppinfo}

The Supporting Information is available free of charge on the ACS Publications website at DOI:

    Computational details and results about topological phase diagram, phonon band structure, molecular dynamics simulations, magnetic ground state, magnetic anisotropy, Tight-binding model, chiral edge states, projected band structures, Berry curvature distribution, and micromagnetic simulations.
\end{suppinfo}

\section{Author Information}
\textbf{Corresponding Authors} \\
*E-mail:xieyu@cumt.edu.cn \\
*E-mail:juntingzhang@cumt.edu.cn \\
\textbf{ORCID} \\
Xinyue Zhu:0009-0002-2244-8685 \\
Yu Xie:0009-0005-2996-1248 \\
Yifei Hao:0009-0002-8394-0657 \\
Fei Gao:0009-0004-3097-4583 \\
Yue Li:0009-0000-2798-6875 \\
Junting Zhang:0000-0003-3268-9685 \\
\textbf{Notes} \\
The authors declare no competing financial interest.

\section{Acknowledgement}
This work was financially supported by the National Natural Science Foundation of China (Grants No. 12374097), and the Fundamental and Interdisciplinary Disciplines Breakthrough Plan of the Ministry of Education of China (Grant No. JYB2025XDXM409).

\bibliography{reference}

\begin{tocentry}
\includegraphics*[width=1\textwidth]{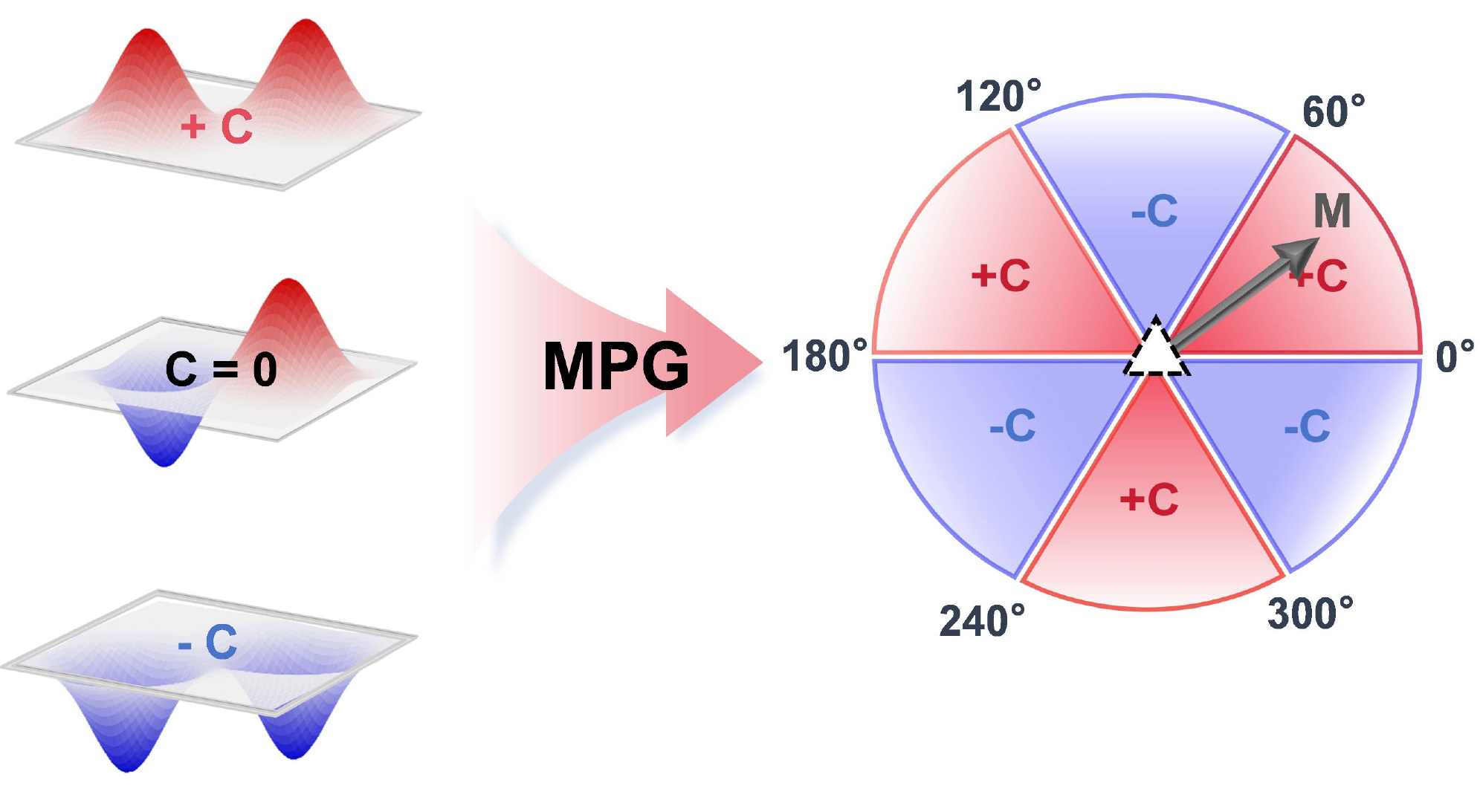}
\end{tocentry}

\end{document}